\documentclass[12pt,preprint]{aastex}

\slugcomment{accepted for publication in ApJ. , 2002}

\shorttitle{Accretion disks around black holes}

\shortauthors{Lee \& Ramirez--Ruiz}

\received{2001 October 11}
\begin{document}

\title{Accretion disks around black holes: dynamical evolution, meridional
circulations and gamma ray bursts}

\author{William H. Lee}
\affil{Instituto de Astronom\'{\i}a, 
Universidad Nacional Aut\'{o}noma de M\'{e}xico, \\ 
Apdo. Postal 70-264, Cd. Universitaria, 
M\'{e}xico D.F. 04510}

\and

\author{Enrico Ramirez--Ruiz} 
\affil{Institute of Astronomy, \\ 
Madingley Road, Cambridge, CB3OHA, U.K.}

\begin{abstract} We study the hydrodynamical evolution of massive
accretion disks around black holes, formed when a neutron star is
disrupted by a black hole in a binary system. The initial conditions
are taken from three--dimensional calculations of coalescing
binaries. By assuming azimuthal symmetry we are able to follow the
time dependence of the disk structure for 0.2~seconds in cylindrical
coordinates $(r,z)$. We use an ideal gas equation of state, and assume
that all the dissipated energy is radiated away. The disks evolve due
to viscous stresses, modeled with an $\alpha$-law. We study the disk
structure, and in particular the strong meridional circulations that
are established and persist throughout our calculations. These consist
of strong outflows along the equatorial plane that reverse direction
close to the surface of the disk and converge on the accretor. In the
context of gamma ray bursts (GRBs), we estimate the energy released
from the system in neutrinos and through magnetic--dominated
mechanisms, and find it can be as high as $E_{\nu}\approx 10^{52}$~erg
and $E_{BZ}\approx 10^{51}$~erg respectively, during an estimated
accretion timescale of 0.1--0.2 seconds. $\nu \overline{\nu}$
annihilation is likely to produce bursts from only a short, impulsive
energy input $L_{\nu \overline{\nu}} \propto t^{-5/2}$ and so would be
unable to account for a large fraction of bursts which show
complicated light curves.  On the other hand, a gas mass $\approx
0.1-0.25 M_\odot$ survives in the orbiting debris, which enables
strong magnetic fields $\approx 10^{16}$~G to be anchored in the dense
matter long enough to power short duration GRBs. We highlight the
effects that the initial disk and black holes masses, viscosity and
binary mass ratio have on the evolution of the disk
structure. Finally, we investigate the continuous energy injection
that arises as the black hole slowly swallows the rest of the disk and
discuss its consequences on the GRB afterglow emission.
\end{abstract}
\keywords{accretion, accretion disks --- hydrodynamics --- gamma rays: bursts}


\section{Introduction}

Accretion onto black holes has been considered as an efficient way to
transform gravitational energy into radiation \citep{s64,z64}, and is
often thought to occur in the form of a disk, due to the angular
momentum of the accreting matter. This almost certainly is the case in
a variety of astrophysical systems, ranging from AGNs with very
massive black holes to stellar mass binaries, analogous to the X--ray
binaries known to contain neutron stars \citep{king95}.  The flows in
these disks may exhibit very different morphologies depending on the
physical conditions present in the system. Accretion is generally
thought to proceed through the transport of angular momentum from the
inner to the outer regions of the disk, although the mechanism by
which this is accomplished is not entirely clear. The parametrization
introduced by \cite{ss73} has allowed much progress to be made,
without specifying the physics behind the viscosity responsible for
angular momentum transport. Magnetohydrodynamical studies (analytical
and numerical) appear to indicate that magnetic fields and their
associated instabilities in disks can effectively generate a viscosity
that would drive their evolution \citep{bh91,hb91,h00,sp01,hk01,hk02},
with equivalent values for the $\alpha$ parameter in the range
0.01--0.1 \citep{bh98}.  Additionally, it has become clear that
hydrodynamic processes can play an important role in the structure and
evolution of disks, and that multi--dimensional, time--dependent
computations are necessary to fully understand these effects,
particularly since the flows can be quite complicated, often
exhibiting strong variability, and composed of combinations of
inflows/outflows of varying intensity \citep{igu00}.

In this context we are motivated to study accretion disks around black
holes hydrodynamically, particularly in what concerns central engines
for cosmological gamma ray bursts (GRBs). The energetics of GRBs
($10^{50}-10^{52}$~erg are typically released in a few seconds) and
the variability shown in their lightcurves (down to millisecond
timescales) argues in favor of a compact source that produces a
relativistic outflow, usually referred to as a fireball
\citep{rm92}. The complicated light curves can then be understood in
terms of internal shocks in the outflow itself, caused by velocity
variations in the expanding plasma
\citep{rm94,sp97,frw99,rf00,np02}. The durations range from
$10^{-3}$~s to about $10^{3}$~s, with a bimodal distribution of short
($\simeq 0.2$~s) and long bursts ($\simeq 40$~s)
\citep{kouveliotou93,n96}.  One possible scenario, at least for the
class of short bursts\footnote{We note, furthermore, that those with
detected afterglows are all in the long category (although see
\cite{lrg01} and \cite{c02}).}, involves the coalescence of compact
binaries containing a black hole (BH) and a neutron star (NS), or two
neutron stars \citep{ls76, bp86, bp91, eich89, nara92, moch93, katz96,
klee98, pop99, rj99, sr02}. Such systems do exist, like PSR1913+16
\citep{ht75}, and they will coalesce due to angular momentum losses to
gravitational radiation, provided that the orbital separation is small
enough. The binary coalescence rates are in rough agreement with the
observed GRB rate \citep{kalo01}. The typical dynamical timescale in
such binaries immediately prior to coalescence ($\approx$~ms) is much
shorter than the observed burst duration, and so it requires that the
central engine evolves into a configuration that is stable, while
retaining a sufficient amount of energy to power the burst.

The formation of a BH with a debris torus around it is a common
ingredient of both these scenarios, whose accretion can provide the
release of sufficient gravitational energy $\approx 10^{54}$ erg to
power a GRB \citep{rees99}. A fireball arises from the large
compressional heating and dissipation associated with this accretion,
which can provide the driving stress necessary for relativistic
expansion \citep{piran99,m01}. Possible forms of this outflow are
kinetic energy of relativistic particles generated by
$\nu\overline{\nu}$ annihilation or an electromagnetic Poynting flux.

In either mechanism, the duration of the burst is determined by the
viscous timescale of the accreting gas\footnote{In the collapsar
scenario, the burst duration is given by the fall-back time of the gas
\citep{w93,mw99}}, which is significantly longer than the dynamical
timescale, thus accounting naturally for the large difference between
the durations of bursts and their fast variability. Any instability
(of hydrodynamical or magnetic origin) would presumably be reflected
in the relativistic outflow. Strong magnetic fields anchored in the
dense matter surrounding the BH would produce large amplitude
variations in the energy release \citep{u92, m97}. A weaker field
would extract inadequate power; on the other hand a neutron torus,
with its huge amount of differential rotation, is a natural site for
the onset of a dynamo process that winds up the magnetic field to the
required intensity \citep{u94, kl98}. An acceptable model requires
that the orbiting debris not be dispersed completely on too short a
timescale, lasting at least as long as the characteristic duration of
the burst.  The tidal disruption of a neutron star by a black hole
would redistribute matter and angular momentum very rapidly. The key
issue is then how long a sufficient amount of this matter survives to
power a burst.\\

In this paper, we study the evolution of realistic disks resulting
from dynamical coalescence calculations (see below \S \ref{method}),
on timescales that are comparable to the durations of short GRBs
(i.e. a few tenths of a second).  Recently, the steady state structure
of similar disks has been examined by \citet{nara01} and
\citet{km02}. Since we wish to investigate the evolution of the disk
and its stability, no steady state assumptions are made regarding its
structure. This allows us to compare the strength of the energy
release by $\nu\overline{\nu}$ annihilation relative to MHD coupling
and to investigate their potential as a viable source for GRB
production.

The hydrodynamics of black hole--neutron star coalescence has been
considered in detail by \citet{lk99a,lk99b} and \citet{l00,l01}, using
an ideal gas equation of state and varying its stiffness through the
adiabatic index $\Gamma$.  The simulations explored the effect of
different initial binary mass ratios $q_{b}=M_{\rm NS}/M_{\rm BH}$ and
the spin configuration of the neutron star, with respect to the
orbital motion. The viscosity inside neutron stars is far too small to
permit synchronization during the inspiral phase \citep{koch92,
bild92}, so it is reasonable to assume that the neutron star spin is
negligible with respect to the orbital angular momentum. The exact
equation of state at supra--nuclear densities is uncertain, but it
would appear that the radius is largely independent of the mass
\citep{prak01}, so that in the polytropic approximation, one would
assume $\Gamma=2$. Coalescence calculations with this set of
parameters were computed by \citet[herafter L01]{l01}. In particular,
we will use here the results of runs C50 and C31 in that paper, with
$q_{b}=0.5$ and $q_{b}=0.31$ respectively (both runs used an index
$\Gamma=2$).

\section{Numerical method and implementation \label{method}}

The equations of hydrodynamics are solved in two dimensions using
Smooth Particle Hydrodynamics (SPH)
\citep{monaghan92}. Three--dimensional dynamical coalescence
calculations are only followed for a few tens of milliseconds, at most
(those in L01 spanned 23~ms), because of computational
limitations. Since we wish to explore the behavior of the accretion
structures on timescales that are much longer than the dynamical one,
we have assumed azimuthal symmetry in the system, and mapped the
output from the 3D calculations to cylindrical $(r,z)$
coordinates. This allows us to follow the calculations for a few
tenths of a second (all the runs shown in this paper were terminated
at $t=0.2$~s). We have not assumed reflection symmetry with respect to
the equatorial ($z=0$) plane.

We use a simple ideal gas equation of state, with $P=\rho u
(\Gamma-1)$ ($u$ is the internal energy per unit mass), and the fluid
is initially given the azimuthal velocity $v_{\phi}$ required for
centrifugal equilibrium. The self--gravity of the disk is neglected,
and the black hole produces a Newtonian point--mass potential
$\Phi_{\rm BH}=-GM_{\rm BH}/r$. Accretion is modeled by an absorbing
boundary at $r_{Sch}=2GM_{\rm BH}/c^{2}$. In practice, only the inner
regions of the 3D disk are mapped onto two dimensions to start the
calculation (with $0<|z|<200$~km and $0<r<400$~km).

The above simplifications clearly do not correspond to the physical
conditions one would encounter in such a disk, but we have chosen to
use them as a first approximation to the solution for several
reasons. In the first place, the 3D dynamical calculations that gave
rise to the disks we now use as initial conditions were fully
Newtonian (except for the inclusion of gravitational radiation
reaction terms), particularly regarding the form of the gravitational
potential produced by the black hole. For consistency we have thus
kept this form. In the future we will explore situations in which the
black hole potential is given by a form which reproduces the effects
of a marginally stable orbit (e.g. that of \citet{pw80}). Second, the
equation of state was likewise very simple in the initial
calculations, and we have maintained it in the present work. The only
change we apply is that at $t=0$ the value of the index $\Gamma$ has
been lowered for some runs (see Table~\ref{table:ICs}) from 2 to
4/3. We will also improve on the form of the equation of state in
future simulations. Third, the final conditions of the 3D calculations
show that the self--gravity of the disk is not too important in
determining its structure (it is pressure support that makes the
rotation curve deviate from what would be expected for point masses
orbiting a massive body, see e.g. Figure~8b in L01).

The second important new ingredient (the first one being the
assumption of azimuthal symmetry) in the present simulations is the
transport of angular momentum by viscosity, which is modeled by
including all the terms derived from the stress tensor $t_{\alpha
\beta}$ in the equations of motion (we use the formalism developed by
\citet{flebbe94}). We have:

\begin{equation}
\frac{dv_{\alpha}}{dt}=-\frac{P_{,\alpha}}{\rho} - \Phi_{\rm
BH,\alpha} + \frac{t_{\alpha \beta, \beta}}{\rho} +
\left(\frac{dv_{\alpha}}{dt}\right)_{art}
\end{equation}
where the commas indicate partial derivatives. The last term arises
from the presence of artificial viscosity (see below). The tensor
responsible for the viscous force is $t_{\alpha \beta}$, given by:
\begin{equation}
t_{\alpha \beta}=\eta (v_{\alpha , \beta}+v_{\beta ,
\alpha}-\frac{2}{3}\delta_{\alpha \beta}v_{\gamma , \gamma}),
\end{equation}
where $\eta$ is the dynamical viscosity coefficient, and the shear is:
\begin{equation}
\sigma_{\alpha \beta}=v_{\alpha , \beta}+v_{\beta ,
\alpha}-\frac{2}{3}\delta_{\alpha \beta}v_{\gamma , \gamma}
\end{equation}
The energy dissipation per unit mass is:
\begin{equation}
T\frac{ds}{dt}=\frac{\eta}{2\rho}\sigma_{\alpha \beta} \sigma_{\alpha
\beta} + \left(T\frac{ds}{dt}\right)_{art} \label {eq:diss}
\end{equation}

To perform a dynamical calculation, at this point one needs to fix a
prescription for the viscosity. We have chosen to parametrize its
magnitude by using an $\alpha$--law, setting $\eta=\alpha \rho
c_{s}^{2}/ \Omega_{k}$, where $c_{s}=(\Gamma P/\rho)^{1/2}$ is the
adiabatic sound speed and $\Omega_{k}$ is the Keplerian angular
velocity. This formulation of the viscosity has been used before in
time--dependent studies of accretion flows in two and three dimensions
\citep{ica96,ia99,ia00,ian00,mg02}. We note that the simulations
reported here are followed for a timescale that is much longer than
the viscous one, and thus solutions that are qualitatively steady are
obtained in all cases (they cannot be steady quantitatively simply
because there is no external agent feeding the disk with matter as the
simulation progresses).

The implementation of SPH for the present 2D azimuthally symmetric
calculations requires the writing of the above expressions in
cylindrical coordinates. For completeness, we give here the full set
of equations for momentum and energy:
\begin{eqnarray}
\frac{dv_{r}}{dt}=-\frac{1}{\rho}\frac{\partial P}{\partial
r}-\frac{GM_{\rm BH}r}{R^{3}}+\frac{1}{\rho}\left(\frac{\partial
t_{rr}}{\partial r}+\frac{t_{rr}}{r}+\frac{\partial t_{rz}}{\partial
z}\right)+\left(\frac{dv_{r}}{dt}\right)_{art}
\end{eqnarray}
\begin{eqnarray}
\frac{dv_{\phi}}{dt}=\frac{1}{\rho}\left(\frac{2t_{r\phi}}{r}+\frac{\partial
t_{\phi r}}{\partial r}+\frac{\partial t_{\phi z}}{\partial z}\right)
\end{eqnarray}
\begin{eqnarray}
\frac{dv_{z}}{dt}=-\frac{1}{\rho}\frac{\partial P}{\partial
z}-\frac{GM_{\rm BH}z}{R^{3}}+\frac{1}{\rho}\left(\frac{\partial
t_{zr}}{\partial r}+\frac{t_{zr}}{r}+\frac{\partial t_{zz}}{\partial
z}\right)+\left(\frac{dv_{z}}{dt}\right)_{art}
\end{eqnarray}
\begin{equation}
\frac{du}{dt}=-\left(\frac{P}{\rho} \right ) \nabla \cdot \vec{v}
+T\frac{ds}{dt}
\end{equation}
Here $R=\sqrt{r^{2}+z^{2}}$ is the distance to the origin, and the
artificial viscosity terms are computed using the prescription of
\citet{bal95} to minimize the effects of artificial shear on the
evolution of the disk. The explicit forms of the acceleration and the
energy dissipation for a given SPH particle are:
\begin{equation}
\left(\frac{d\vec{v}}{dt}\right)_{i,art}=-\sum_{j \neq i}m_{j}\Pi_{ij}
\nabla_{i} W_{ij},
\end{equation}
and
\begin{eqnarray}
\left(T\frac{ds}{dt}\right)_{i,art}=\frac{1}{2}\sum_{j\neq i}m_{j}
\Pi_{ij}(\vec{v}_{i}-\vec{v}_{j}) \cdot \nabla_{i} W_{ij},
\end{eqnarray}
where $W$ is the SPH interpolation kernel given by \citet{ml85},
adapted to azimuthal symmetry:
\begin{eqnarray}
W(r,h) = \frac{10}{7 \pi h^{2}} \frac{1}{(2\pi r)}\left\{ \begin{array}{ll} 1-3
		    \left( r/h \right)^{2}/2+3 \left(
		    r/h \right) ^{3}/4, & 0 \leq r/h < 1, \\
		    \left(2-r/h \right) ^{3}/4, & 1
		    \leq r/h < 2, \\ 0, & 2 \leq r/h.  \end{array}
		    \right. \label{eq:defW}
\end{eqnarray}
We have
\begin{eqnarray}
\Pi_{ij}=\left( \frac{P_{i}}{\rho^{2}_{i}} +
\frac{P_{j}}{\rho^{2}_{j}}\right)(-\alpha_{b} \mu_{ij} +\beta_{b}
\mu^{2}_{ij}),
\end{eqnarray} 
where
\begin{eqnarray*}
\mu_{ij}=\left \{ \begin{array}{ll} \frac{(\mbox{\boldmath
$v$}_{i}-\mbox{\boldmath $v$}_{j}) \cdot (\mbox{\boldmath
$r$}_{i}-\mbox{\boldmath $r$}_{j})}{h_{ij}(|\mbox{\boldmath
$r$}_{i}-\mbox{\boldmath $r$}_{j}|^{2}/h_{ij}^{2})+\eta^{2}}
\frac{f_{i}+f_{j}}{2c_{ij}}, & (\mbox{\boldmath
$v$}_{i}-\mbox{\boldmath $v$}_{j}) \cdot (\mbox{\boldmath
$r$}_{i}-\mbox{\boldmath $r$}_{j}) <0 \\ 0, & (\mbox{\boldmath
$v$}_{i}-\mbox{\boldmath $v$}_{j}) \cdot (\mbox{\boldmath
$r$}_{i}-\mbox{\boldmath $r$}_{j}) \geq 0
\end{array} \right.
\end{eqnarray*}
and $f_{i}$ is the form-function for particle {\em i} defined by
\begin{eqnarray*}
f_{i}=\frac{|\mbox{\boldmath $\nabla$} \cdot \mbox{\boldmath
$v$}|_{i}}{|\mbox{\boldmath $\nabla$} \cdot \mbox{\boldmath
$v$}|_{i}+|\mbox{\boldmath $\nabla$} \times \mbox{\boldmath
$v$}|_{i}+\eta'c_{i}/h_{i}}.
\end{eqnarray*}
The factor $\eta'\simeq 10^{-4}$ in the denominator prevents numerical
divergences. The smoothing length and sound speed for particle {\em i}
are denoted by $h_{i}$ and $c_{i}$ respectively, and $\alpha_{b}$ and
$\beta_{b}$ are constants of order unity (we use
$\alpha_{b}=\beta_{b}=\Gamma/2$). The number of neighbors for the
hydrodynamical interpolation is kept fixed at $\nu=24$.

Artificial viscosity is used in SPH to model the presence of shocks
and avoid particle inter-penetration. The forms initially employed
have been substantially improved over the years, with an accompanying
reduction in spurious numerical effects. Most notably, the one we use
here is such that the shear viscosity is suppressed when the
compression in the fluid is low and the vorticity is high ($|\nabla
\times \vec{v}| \gg |\nabla \cdot \vec{v}|$, as is the case in
accretion disks with differential rotation), but remains in effect if
compression dominates in the flow ($|\nabla \cdot \vec{v}| \gg |\nabla
\times \vec{v}|$).

We are able to monitor just how important the dissipation due to
artificial viscosity is, compared with that arising from the stress
tensor $t_{\alpha \beta}$ (see equation~\ref{eq:diss}). We employed
two different values of the viscosity parameter $\alpha$ for the runs
shown here (see Table~\ref{table:ICs}):$\alpha=0.1$ and
$\alpha=0.01$. For both cases, the energy dissipated through
artificial viscosity is most important at early times ($t<10$~ms), and
quickly decreases relative to the terms generated through $t_{\alpha
\beta}$ (see below, \S \ref{results}).

We have tested our code to ensure that the viscous terms are computed
correctly. The most simple test is to calculate the spreading of a
ring of gas due to viscosity (with no vertical structure) orbiting a
point mass, when pressure effects and self--gravity are
neglected. This problem has an analytic solution in terms of Bessel
functions \citep{pringle81}, which can be easily compared with a
numerical solution. Our code reproduces this calculation accurately,
with the ring spreading radially on the correct timescale.

We have also performed a second class of tests, allowing for vertical
structure in azimuthal symmetry, studying the evolution of thick tori,
where pressure effects are important and self--gravity is
neglected. Two different types of calculations were performed. In the
first, a torus in hydrostatic equilibrium around a black hole, with an
initially specified distribution of specific angular momentum, was
allowed to evolve, switching the viscosity on at the start of the
calculation \citep{ica96}. In the second, a continuous flow of matter
was injected at a large radius, and a disk was formed on the viscous
timescale \citep{ia99}. For each of these, several values of the
$\alpha$ parameter were used, ranging from $10^{-3}$ to $10^{-1}$. We
found excellent agreement with previously published results, both
qualitatively and quantitatively \citep{lmuqso3}.

Finally, we assume that all the energy dissipated in the flow is
radiated away in neutrinos, and thus in this respect, the flow is an
NDAF (Neutrino Dominated Accretion Flow), as done by \citet{nara01}
and \citet{km02}. This is accomplished by directly removing from the
fluid the internal energy associated with the first term in equation
(\ref{eq:diss}). The neutrino luminosity is then given by
$L_{\nu}=\int \eta \sigma_{\alpha \beta} \sigma_{\alpha \beta} /2 \rho
\ dm$. The energy dissipated by artificial viscosity remains in the disk
as internal energy.  This is another aspect that clearly needs to be
treated in a more realistic manner and will be dealt with in future
work.

\section{Results \label{results}}

\subsection{Structure and fluid circulation in the disks \label{circulation}}

The choice of parameters shown in Table~\ref{table:ICs} for the
simulations allows us to explore the effect of numerical resolution on
the results (runs D and E, and runs F and G differ only in the initial
number of particles), as well as that of the magnitude of the
viscosity, the compressibility of the gas in the disk, and the initial
mass of the black hole. For a lower black hole mass $M_{\rm BH}$ the
accretion disk is initially more massive, and the spin angular
momentum of the black hole is larger. This is a reflection of the
previous binary evolution and tidal disruption that led to the
formation of the accretion disk.

The initial conditions we have used for the accretion disks clearly do
not correspond to a simple equilibrium configuration, since they are
derived from three--dimensional dynamical calculations. In our case,
the disks are initially hot (the internal energy is about
10~MeV/nucleon, see Figure~8 in L01), and thus pressure support is
important, making them quite thick ($H/R\approx 0.5$). This energy
reservoir is not entirely lost, since only the energy dissipated
through the physical viscosity is radiated away in neutrinos. Thus the
disks remain thick throughout the dynamical evolution.

All the dynamical calculations are qualitatively similar. We show
snapshots of the density and velocity field in the disk at various
times for runs E ($\alpha=0.1$) and G ($\alpha=0.01$) in
Figures~\ref{rhoEG} and \ref{vEG}. At the start of the calculation,
the disk flattens slightly, on a vertical free--fall timescale
$t_{ff}\approx\sqrt{r^{3}/GM_{\rm BH}}\approx 1-2$~ms, since it is not
in strict hydrostatic equilibrium at $t=0$. The initial panels (at
$t=10$~ms) show the formation of a weak shock at $r\simeq 140$~km,
formed when the outer regions of the torus move radially inwards. This
front moves subsequently outwards, and has practically left the inner
region of the disk by $t=20$~ms. Thereafter accretion proceeds onto
the black hole in a characteristic pattern, which is mainly dependent
on the value of $\alpha$.

The simulations use a variable smoothing length $h$ to ensure an
accurate interpolation of the fluid variables, and so the spatial
resolution (which is essentially equal to $h$), is also variable. It
is smallest in the regions of highest density (in the midplane of the
disk). At $t=0.1$~s, the smallest scales that can be resolved are of
order $2 \times 10^{4}$~cm=200~m at $r\simeq 30$~km, and $3 \times
10^{5}$~cm=3~km close to the surface of the disk (where $\rho \simeq
10^{9}$~g~cm$^{-3}$ for the high--resolution runs E and G; in two
dimensions the smoothing length scales with the number of particles as
$h \propto N^{-1/2}$). The circulation patterns seen in
Figure~\ref{vEG} are much larger than the spatial resolution and thus
are clearly resolved. The runs with a lower number of particles have a
slightly poorer resolution, but nevertheless resolve the same global
features clearly in terms of accretion rate, spatial extension of the
disk and circulation patterns.

For a high viscosity, a large--scale circulation is established, with
$z\approx r$. There is outflow in the equatorial plane at large
distances from the black hole, and accretion takes place only along
the surface of the disk and in the equatorial plane quite close to the
black hole ($r\simeq 50$~km). The radial velocities are subsonic, and
the turnaround in $v_{r}$ occurs in large eddies close to the surface
of the disk, when the gas is moving away from the equatorial
plane. The disk slowly spreads in the radial direction, with an
accompanying drop in density and accretion rate (at $t=0.2$~s,
$\dot{M}$ has dropped by about one order of magnitude compared to its
value at 10~ms, see also Figure~\ref{mdotbz}a below).

For the low viscosity case, the flow is more complicated. Accretion
onto the black hole still occurs only along the surface of the disk
and in the equatorial plane at small values of $r$. However, the
circulation patterns are present on smaller scales than before, with
$z<r$. The flow is qualitatively steady after approximately 20~ms, but
the eddies persist throughout the simulation. The radial spread of the
disk is less pronounced than before, due to the lower value of
$\alpha$, and thus higher densities are maintained for a longer
period. The accretion rate onto the black hole is about one order of
magnitude lower (for $t<20$~ms) than for the high viscosity runs, and
decreases more slowly.
 
The circulation patterns described above can be clearly related to the
distribution of specific angular momentum $l$ in the disk. We show in
Figure~\ref{lrz} the contours of $l$ for runs E and G at $t=40$~ms. As
one moves from the equator (where $v_{r}>0)$ to the surface of the
disk at a given value of $r$, $l$ decreases until the fluid turns
around and flows inward.

For both cases, the magnitude of the velocities (radial and vertical)
in the disk decreases in time, with the circulation patterns becoming
less intense in that respect. Figure~\ref{LS} shows views of the disks
for runs E and G at $t=0.11$~s on a larger scale.

The mass accretion rate onto the black hole is clearly sensitive to
the value of $\alpha$, and to a lesser extent, on the adiabatic index
$\Gamma$, as can be seen in Figure~\ref{mdotbz}a. Runs E and G differ
only in the value of $\alpha$, and clearly lowering the viscosity
leads to less vigorous transport of angular momentum, and thus to a
reduced $\dot{M}_{\rm BH}$. Runs E and C differ only in the value of
$\Gamma$. The less compressible case (run C, with $\Gamma=2$) has a
broader spatial mass distribution and reaches lower peak
densities. Thus slightly more mass is initially closer to the accretor
and easily absorbed, giving a larger $\dot{M}_{\rm BH}$, and
subsequently a faster drop (by $t=0.2$~s, the disk mass is lower in
run C than in run E, see Table~\ref{table:evol}). Finally, run B
differs from run E in the initial binary system that gave rise to the
accretion disk, with a larger $M_{\rm BH}$ and smaller $M_{disk}$,
which is responsible for the lower accretion rate.

Meridional circulation patterns like the ones seen in the present
simulations have been studied before, both analytically and
numerically. Solving for the vertical structure of an accretion disk
and using an $\alpha$ viscosity, \cite{urpin84} found that the radial
flow can change direction in the midplane of the disk, with $v_{r}>0$.
\cite{kita95} and \cite{kk00} assumed a steady state in a polytropic
$\alpha$--disk and found solutions which exhibited outflow along the
equator, with inflow along the disk surface and in the equator close
to the accretor, as seen here. In their study it was the value of
$\alpha$ that fixed the equatorial distance at which the flow turned
from inflow to outflow (the stagnation radius). In both cases, the
authors assumed that the disk cools efficiently, with the dissipated
energy being radiated away.

To our knowledge, numerical work has been performed so far mainly with
Eulerian codes, in which boundary conditions need to be
specified. \cite{kl92} obtained outflow in the midplane of the disk,
and inflow along the surface, as in the solution found by
\cite{urpin84}, but the use of height--averaged boundary conditions
enforced artificial circulation at the edges of the computational
domain, thus altering the flow structure substantially.  In other
cases, a free outflow condition is used, meaning that gas which leaves
the computational domain cannot re--enter it at a later time. These
calculations --- which, contrary to what we have assumed, suppose that
the dissipated energy is advected with the flow --- have also revealed
circulation patterns and inflows/outflows with morphologies that
depend on the magnitude of the viscosity \citep{ia99,spb99,mg02}.

The use of SPH is clearly an advantage in this sense, since there are
no boundary conditions at all and the modeling of the fluid is not
restricted to a particular region in space.

\subsection{Dynamical stability \label{stability}}

There are several instabilities that can affect massive accretion
disks dynamically and shorten their lifetimes considerably, compared
with the viscous timescale. One of these is the ``runaway radial''
instability, first discussed in the context of massive accretion disks
at the centers of galaxies \citep{abra83}. A number of different
effects can work to enhance or suppress this phenomenon, namely (i)
the spin of the black hole, (ii) the distribution of angular momentum
in the disk, given by $l\propto r^{p}$, (iii) the effects of general
relativity and (iv) the self--gravity of the disk. The first two
(large Kerr parameter and a high value of $p$) tend to suppress the
instability \citep{wilson84,daigne,abram98,masuda98,lu00}, while the
others tend to enhance it
\citep{nishida96a,nishida96b,masuda98,fd02}. The numerical approach
used here allows us only to comment on (ii).  The distribution of
angular momentum in the disk can be approximated by a power law in the
three--dimensional SPH coalescence simulations, with $l\propto
r^{0.45}$, i.e. quite close to Keplerian (see Figures 10b and 8b in
\citet{l00} and L01 respectively). This does not vary appreciably as
the calculations progress, despite the circulation of the fluid in the
disk described above in \S \ref{circulation}, and thus we find that
the disk is stable in this respect.

Additionally, disks can be unstable to axisymmetric perturbations
depending on their surface density and temperature (measured through
the sound speed $c_{s}$). This is known as Toomre's stability
criterion, and can be quantified through $Q=c_{s}\kappa/ \pi G\Sigma$,
where $\kappa$ is the local epicyclic frequency and $\Sigma$ is the
mass surface density:
\begin{equation}
\Sigma(r)=\int_{-\infty}^{\infty} \rho(r,z) \ dz.
\end{equation}
For $Q>1$ the disk is stable to axisymmetric perturbations, and
unstable otherwise. This condition tells us that a dense, cold disk is
more likely to be unstable than a hot, thick one (such as the ones
treated here). We find that for all our runs, $Q>1$. The profile of
$Q$ is such that at a given instant in time, there is a minimum
$Q_{min}$ at a certain radius (close to the maximum in density, see
Figure~\ref{Qr}). The lowest value of $Q_{min}$ occurs at $t=0$, where
$(r_{Q},Q_{min})\approx(50{\rm ~km},5)$. As the disk evolves, the
density drops and both parameters increase. By $t=0.2$~s, we typically
find $(r_{Q},Q_{min})\approx(150{\rm ~km},10)$.

\subsection{Relative importance of artificial viscosity \label{artificial}}

The artificial viscosity mentioned above in \S \ref{method} ideally
should not affect the accretion flow substantially, otherwise the
results would be difficult to assess. As a measure of the effect the
terms arising from it have on the disk, we plot in Figure~\ref{lart}
the dissipated energy as a function of time, separated into the
components arising from $t_{\alpha \beta}$ and those coming from
artificial viscosity, for runs E and G (the former is in fact what we
have termed the neutrino luminosity $L_{\nu}$ above in \S
\ref{method}). It is at early times ($t<10$~ms) that artificial
viscosity is important. This is not surprising, since that is when the
initial transient appears in the disk and the shock forms at $r\approx
140$~km. Thus the relative importance of the artificial viscosity
merely means that it is working as one would expect. After this
transient dies out, the dissipation rate becomes quickly dominated by
the physical viscosity. It is also apparent that only for low values
of $\alpha$ (run G with $\alpha=0.01$) are the two terms comparable in
magnitude in the initial stages. We are thus confident that our
results are not affected significantly by the presence of artificial
viscosity.

\section{Summary and discussion}

We have computed the dynamical evolution of massive accretion disks
around stellar-mass black holes in two dimensions (with azimuthal
symmetry), formed through the tidal disruption of a neutron star by a
black hole in a close binary ($M_{\rm BH}\approx 4M_{\sun}$ and
$M_{disk}\approx 0.3M_{\sun}$). Our initial conditions are taken from
the final state of three dimensional hydrodynamical calculations of
the coalescence process, by averaging in the azimuthal direction. We
use Newtonian physics, an ideal gas equation of state, and solve the
equations of viscous hydrodynamics assuming an $\alpha$ law for the
viscosity coefficient. All the energy dissipated by the physical
viscosity is radiated away (in neutrinos). The time evolution is
followed for 0.2~seconds.

We find that after an initial transient of numerical origin, stemming
from the fact that the 3D torus does not exhibit strict azimuthal
symmetry due to the highly dynamical merging process (see Figures 2
and 7 in L01), the disk settles to a qualitatively steady
state. Meridional circulations are promptly established, whose
structure depends mainly on the value of the $\alpha$ parameter. Most
strikingly, there is an important motion of fluid from the inner
regions of the disks to large radii, along the equatorial plane. The
flow is directed towards the accreting black hole along the surface of
the disk and in the equatorial region at small radii. The disks remain
thick ($H/R \simeq 0.5$) throughout the dynamical evolution, due to
their large internal energy, with accretion rates on the order of one
solar mass per second. The maximum densities decrease during our
calculations, as there is no external agent feeding the disks, but
remain at $\simeq 10^{12}$~g~cm$^{-3}$, with corresponding internal
energy densities $\simeq {\rm few} \times 10^{30}$~erg~cm$^{-3}$.

We stress that the evolution of accretion disks such as these should
be studied with time--dependent models, since the system is clearly
not in a steady state, even from its inception. The circulation
pattern seen in Figure \ref{vEG} would certainly lengthen their
lifetimes by moving matter to larger radii continuously, an effect
that would otherwise be omitted. To illustrate how important this can
be, we considered the total radial mass flux in the disk, composed of
two parts at any given value of the radial coordinate $r$,
$\dot{M}(r)=\dot{M}_{in}(r)+\dot{M}_{out}(r)$ where
\begin{equation}
\dot{M}_{in}=2 \pi r \int_{v_{r}<0} \rho v_{r} dz, \; \;
\dot{M}_{out}=2 \pi r \int_{v_{r}>0} \rho v_{r} dz,
\end{equation}
restricted to regions in which $v_{r}<0$ for $\dot{M}_{in}$ and
$v_{r}>0$ for $\dot{M}_{out}$ respectively. By definition,
$\dot{M}_{in}<0$ and $\dot{M}_{out}>0$. If there were no circulation
in the disk and all matter moved radially inward one would have
$\dot{M}_{out}=0$ and $\dot{M}=\dot{M}_{in}<0$. A measure of how
important the circulations are, and how much they would lengthen the
lifetime of the disks can be obtained by calculating the fraction of
the gas flowing radially in the disk that is actually moving toward
the accretor,
i.e. $|\dot{M}_{in}|/(|\dot{M}_{in}|+|\dot{M}_{out}|)$. In the inner
regions of the disks this ratio tends to unity, as can be seen from
Figure~\ref{vEG}. It decreases rapidly at larger radii, reaching about
1/3 for run E and 1/10 for run G at $r \simeq 100$~km midway through
the simulations (at $t=0.1$~s). Thus most of the radial flow of gas
actually cancels out in the circulations, with a residual amount left
over moving toward the black hole.

We now turn to the implications our calculations might have on models
for the production of cosmological gamma ray bursts from coalescing
compact binaries. The two main forms of energy release from the disk
we consider are i) neutrino emission and ii) MHD flow, either through
the Blandford--Znajek effect \citep{bz77} or by means of a magnetized
wind.

\begin{figure}
\plotone{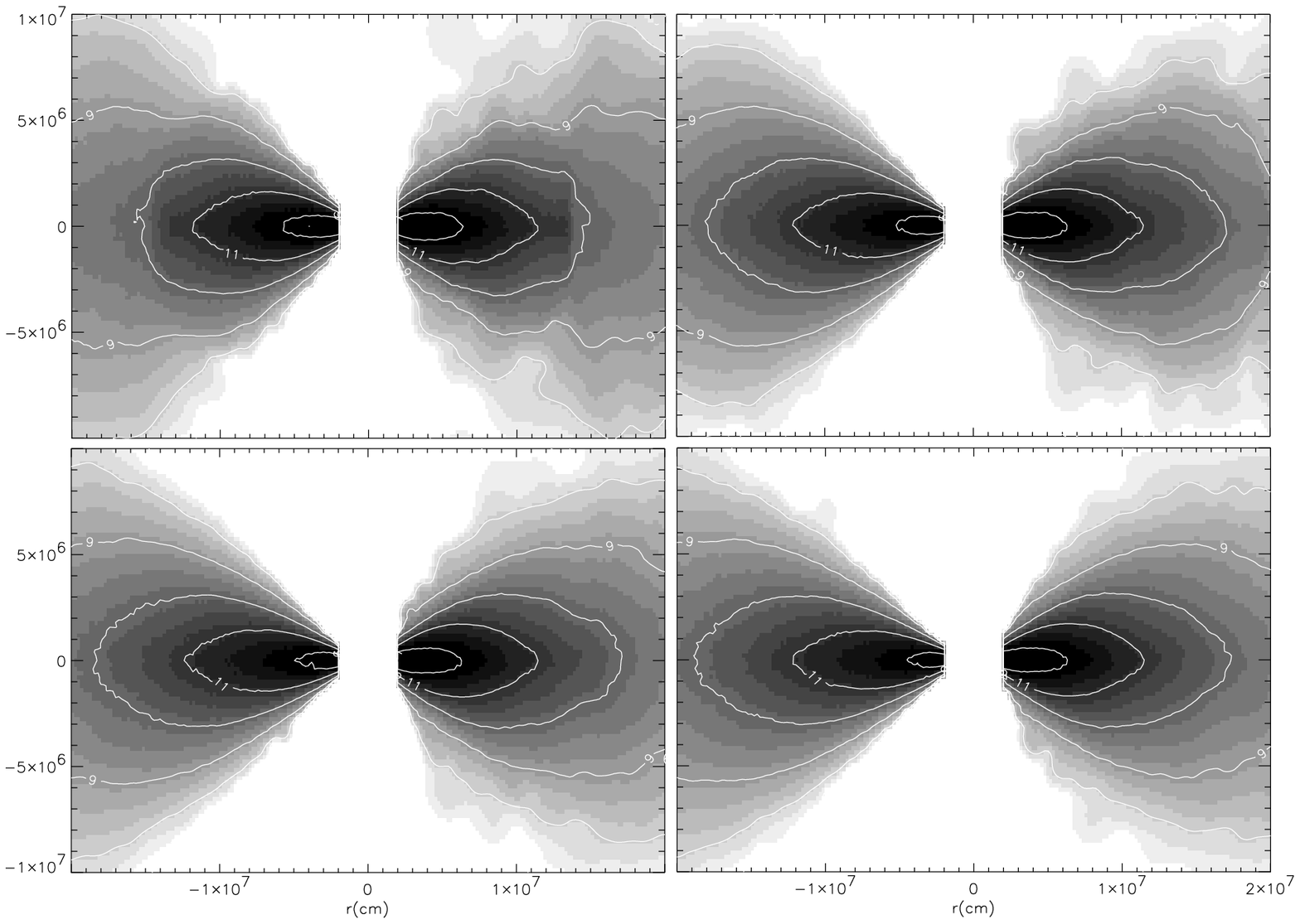}
\caption{Logarithmic density contours (equally spaced and labeled in
g~cm$^{-3}$) at $t=10$~ms (top left), $t=20$~ms (top right), $t=30$~ms
(bottom left) and $t=40$~ms (bottom right) for runs E and G. In each
panel, the left half ($r<0$) corresponds to run E ($\alpha=0.1$) while
the right half ($r>0$) corresponds to run G ($\alpha=0.01$).
\label{rhoEG}}
\end{figure}

\begin{figure}
\plotone{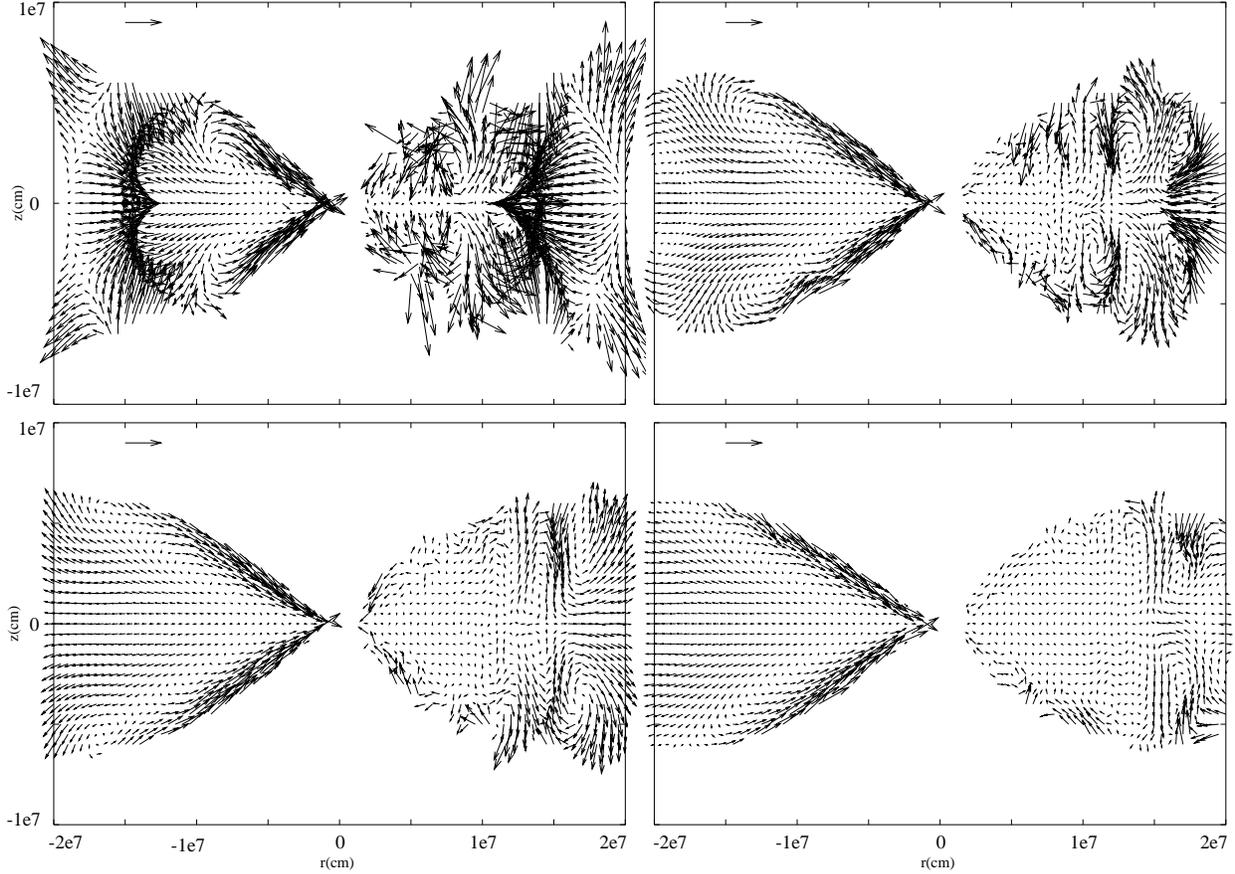}
\caption{Same as Figure~\ref{rhoEG} but showing the velocity
field. The equatorial outflow at large radii, and the inflow along the
surface of the disk can be clearly seen, particularly for run E (left
half of each panel). Note the small--scale eddies present in run G in
every panel. The strength of the circulation diminishes in time, but
does not die out. The velocity field is only shown if $\rho>6 \times
10^{8}$~g~cm$^{-3}$ for clarity (in these panels the edge of the disk,
as given by the SPH particle distribution extends to lower densities,
by about one order of magnitude, see Figure~\ref{rhoEG}). The vector
in the top left corner of each panel has magnitude $v=5 \times
10^{8}$~cm~s$^{-1}$.
\label{vEG}}
\end{figure}

\begin{figure}
\plotone{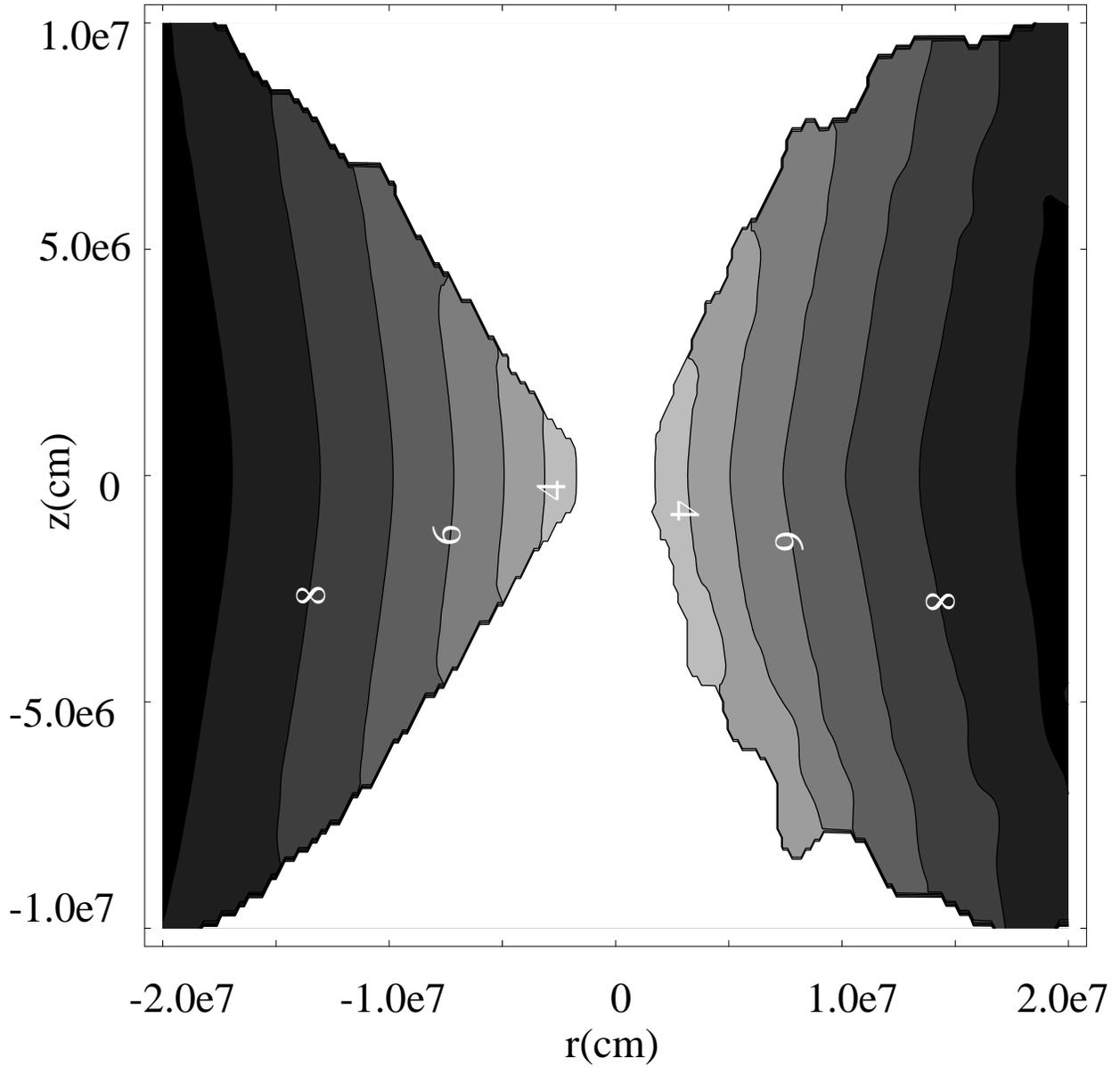}
\caption{Contours of specific angular momentum $l$ for runs E
($\alpha=0.1$) and G ($\alpha=0.01$) at $t=40$~ms (corresponding to
the last panel in Figures~\ref{rhoEG} and \ref{vEG}). The left half of
the panel ($r<0$) corresponds to run E, and the right half ($r>0$) to
run G. The contours are equally spaced and the labels correspond to
$l/10^{16}$~cm$^{2}$~s$^{-1}$.}
\label{lrz}
\end{figure}

\begin{figure}
\epsscale{0.8}
\plotone{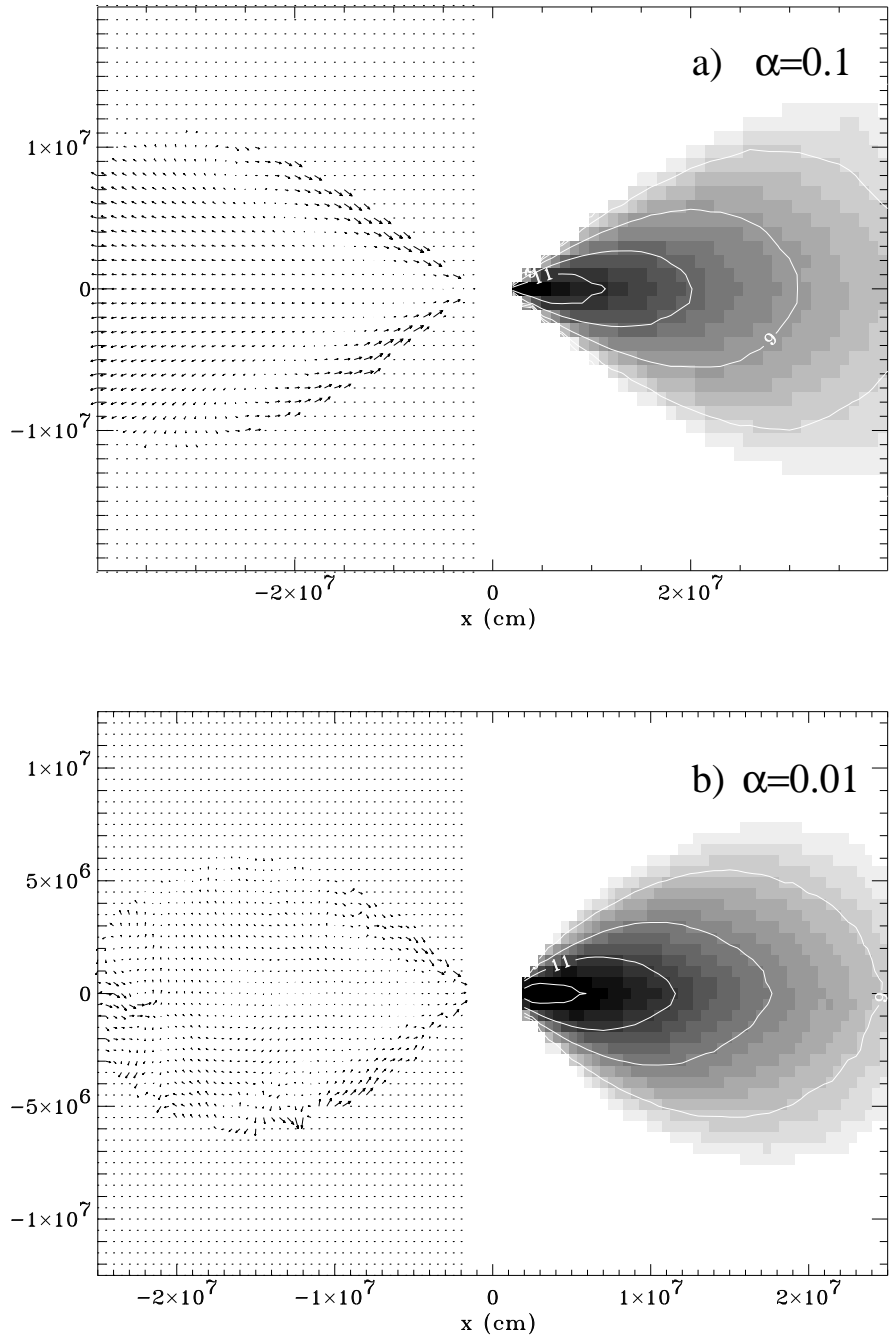}
\caption{Logarithmic density contours (equally spaced and labeled in
g~cm$^{-3}$), and velocity field for run E (a) and G (b) at
$t=0.11$~s. Note the different scales on the axes for each panel. The
longest arrows correspond to $v=2.9 \times^{8}$cm~s$^{-1}$ for (a) and
$v=  10^{8}$~cm~s$^{-1}$ for (b).
\label{LS}}
\end{figure}

\begin{figure}
\plotone{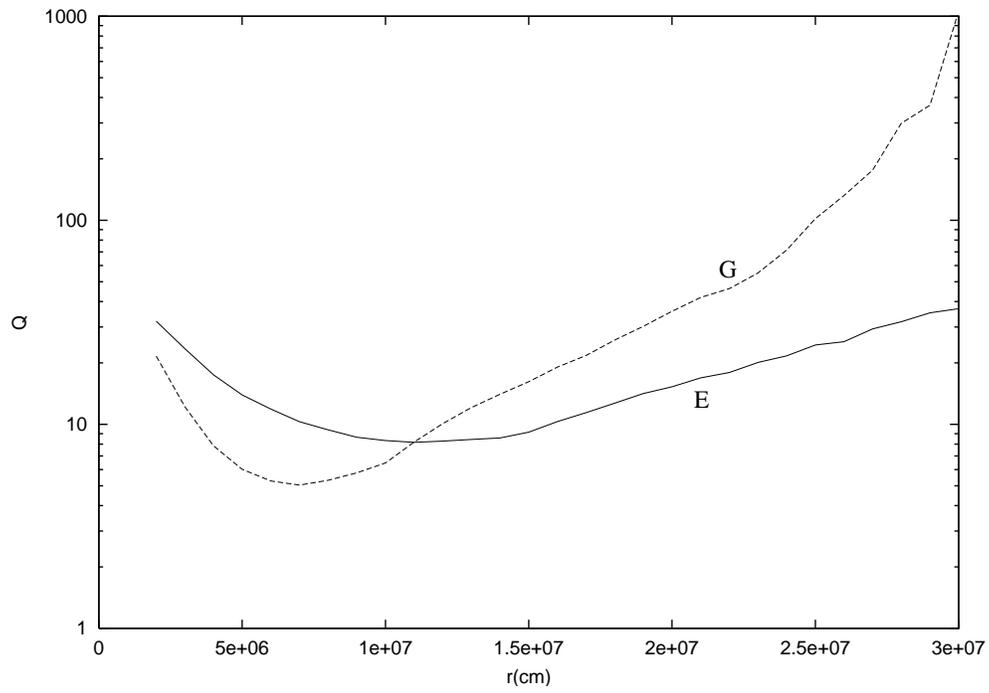}
\caption{Toomre stability parameter $Q$ as a function of the radial
coordinate $r$ for runs E (solid line, $\alpha=0.1$) and G (dashed
line, $\alpha=0.01$), at $t=0.1$~s. }
\label{Qr}
\end{figure}

\begin{figure}
\plotone{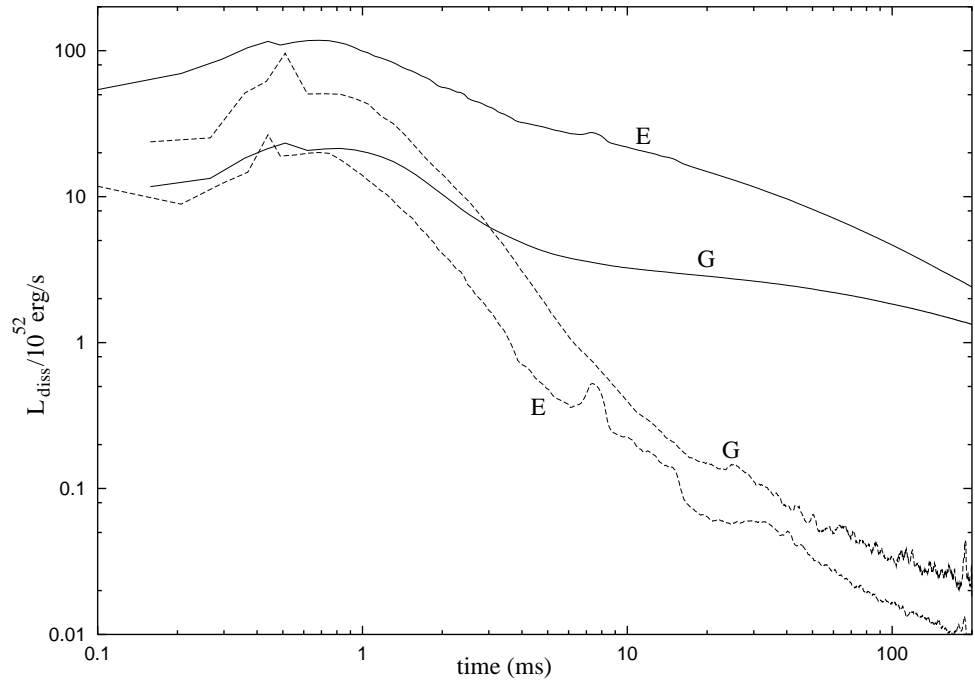}
\caption{Energy dissipation rate as a function of time for runs E and
G arising from the stress tensor $t_{\alpha \beta}$ (solid lines,
$L_{\nu}$), and from artificial viscosity (dashed lines).}
\label{lart}
\end{figure}

\begin{figure}
\epsscale{1.2}
\plotone{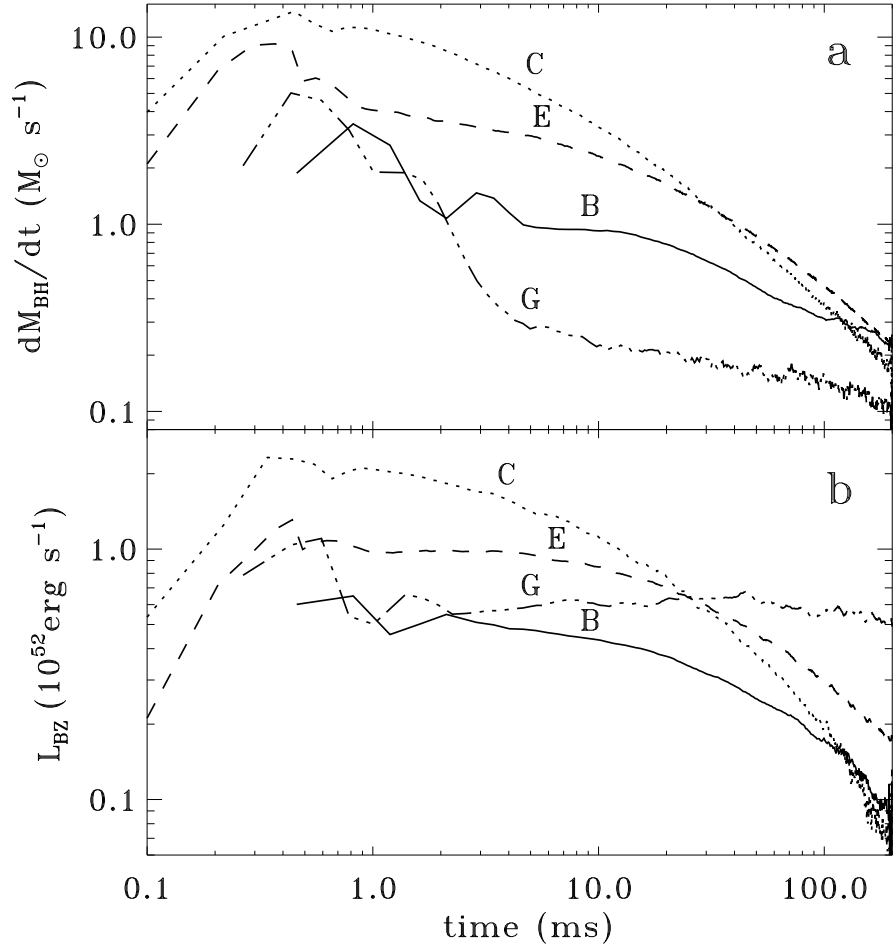}
\caption{(a) Accretion rate (in solar masses per second) onto the
black hole for runs B, C, E and G. (b) Blandford--Znajek luminosity
$L_{BZ}$ for the same runs as shown in (a).
\label{mdotbz}}
\end{figure}

In the first case, we make a rough estimate for $L_{\nu}$ and
$E_{\nu}$ from the dissipated energy because of viscosity (see
Table~\ref{table:evol}), as mentioned above.  We have made an estimate
only for the total neutrino luminosity $L_{\nu}$, and not for the {\em
annihilation} luminosity $L_{\nu \overline{\nu}}$, which would
determine if a relativistic fireball could be launched or not. The
calculation of $L_{\nu \overline{\nu}}$ requires the use of a more
realistic equation of state, which we will explore in future work. For
the time being, we note that, regardless of the efficiency of energy
conversion from neutrino luminosity to annihilation luminosity ---
which could be quite low, on the order of 1 per cent or less
\citep{rj99,pop99}--- it appears that the time--dependence of
$L_{\nu}$, which follows that of $\dot{M}_{\rm BH}$ (see Figure
\ref{mdotbz}a and Table \ref{table:evol}), is such that neutrinos
could only be responsible for a very short, almost impulsive energy
release ($L_{\nu}\propto t^{-5/4}$, so $L_{\nu \overline{\nu}} \propto
L_{\nu}^{2}\propto t^{-5/2}$), and thus would be unable to power a
burst lasting several tenths of a second or more. Of course this does
not mean that it would have a negligible impact on the structure of
the burst itself. We note furthermore that in the detailed 3D
calculations done by \cite{rj99} for the evolution of thick disks
following the coalescence of two neutron stars, the ``neutrinosphere''
is quite close to an isodensity surface at $\rho=10^{11}$~g~cm$^{-3}$,
which is lower than the maximum densities present in the disks
computed here, even at late times (see Figures~\ref{rhoEG} and
\ref{LS}). \cite{sr02} have also performed 3D calculations of binary
neutron star coalescence taking into account neutrino emission and
scattering processes, finding as well that in the regions of highest
density the material is opaque, in their case mainly because of
scattering off heavy nuclei. Thus it is clear that a complete picture
must include an appropriate formulation of neutrino transport.

For the magnetic--dominated case we must make some assumptions, since
our simulations do not incorporate the effects of magnetic fields
explicitly. For the field to be able to extract the binding energy of
the torus, it should be anchored to it, and we assume its magnitude is
directly related to the internal energy in the gas. For this purpose
we show in Table~\ref{table:evol} the internal energy density $\rho
c_{s}^{2}$ at $t=0.1$~s in the inner regions of the disk (at $r=1.25
r_{Sch}$ ($\approx 15$~km for runs A and B, and $\approx 20$~km for
runs C through G) in the equatorial plane. It is of order
$10^{30}$~erg~cm$^{-3}$, and even larger for the case with low
viscosity ($\alpha=0.01$). From this we compute an estimate for the
Blandford--Znajek luminosity as
\begin{equation}
L_{BZ}\approx10^{50} a^{2} \left(\frac{M_{\rm
BH}}{3~M_{\sun}}\right)^{2} \left(\frac{B}{10^{15}~{\rm
G}}\right)^{2}~{\rm erg}~{\rm s}^{-1}
\end{equation} 
where $a \simeq 0.3$ is the Kerr parameter of the black hole and the
magnitude of the magnetic field is computed using $B^{2}/8\pi=\rho
c_{s}^{2}$ (this gives $B\approx 10^{16}$~G in all cases). This is
clearly the most optimistic scenario concerning energy release, in
that it assumes that the magnetic field strength is at the
equipartition value. The time evolution of $L_{BZ}$ is shown in
Figure~\ref{mdotbz}b for several runs.  For a larger viscosity, the
gas in the disk drains into the black hole on a shorter timescale, and
thus the drop in $L_{BZ}$ is much faster than for a low value of
$\alpha$ (in run G, $L_{BZ}\approx 5 \times 10^{51}$~erg~s$^{-1}$ is
practically constant).

In principle, energy can be extracted over many dynamical timescales
if magnetic fields can tap the rotational energy of the accretion disk
and of the black hole. Here $L_{BZ} \propto (t/t_0)^{q^{\prime}}$ is
the intrinsic luminosity of the central engine measured in the fixed
frame, where $-5/4<q^{\prime}<0$ (see Figure \ref{mdotbz}b). For such
a continuously fed fireball, a forward shock propagating into the
external medium and a reverse shock moving back into the relativistic
outflow will coexist on either side of the contact discontinuity. The
latter may persist as long as a significant level of energy injection
is maintained. The differential energy conservation relation for the
self-similar blast wave can be written as $dE/dt
=L_{BZ}(t/t_0)^{q^{\prime}} -\kappa^{\prime}(E/t)$ \citep{cps98},
where the first term denotes steady energy injection, and the second
accounts for possible radiative energy losses of the fireball, with
$q^{\prime}$ and $\kappa^{\prime}$ being dimensionless constants. In
the adiabatic case (no radiative losses) $\kappa^{\prime}=0$ and the
Lorentz factor evolves as $\Gamma^2 \propto t^{-3}$ \citep{bm76}.  For
$q^{\prime}\neq -1-\kappa^{\prime}$, an analytical solution can be
found: $E=[L_{BZ}/(q^{\prime}+\kappa^{\prime} +1)](t/t_0)^{q^{\prime}}
+ E_{\rm imp}(t/t_0)^{-\kappa^{\prime}}$ for $t>t_0$, where $E_{\rm
imp}=E_{\nu \overline{\nu}}$ describes the impulsive energy input.
Here $t_0$ is the characteristic timescale for the formation of a
self-similar solution, which is roughly equal to the time for the
external shock to start to decelerate. For $t>t_0$, the bulk Lorentz
factor of the fireball scales with time as $\Gamma^2 \propto t^{-m}$,
with $m$ and $\kappa^{\prime}$ connected by $\kappa^{\prime}=m-3$, and
$m=3$ for an adiabatic blast wave expanding in a constant density
medium. In the observer frame, the arrival time at the detector $T$ is
related to that in the fixed (laboratory) frame $t$ by
$dT=(1-\beta)dt$, and $T=\int_0^{t} (2\Gamma^2)^{-1}dt \approx t/[2(m
+1)\Gamma^2]$ \citep{fcrs99}.  The differential energy conservation
relation in the observer frame is now given by $dE/dT
={\cal{L}}_{BZ}(T/T_0)^{q} -\kappa(E/t)$, and the integrated relation
is
\begin{equation}
E={{\cal{L}}_{BZ} \over \kappa+q+1} \left({T \over T_0}\right)^q T +
E_{\nu \overline{\nu}} \left({T \over T_0}\right)^{-\kappa},\;\;\; T >
T_0.
\end{equation} 
Here ${\cal{L}}_{BZ}=2\Gamma^2(t_0)L_{BZ}$, and
$q=(q^{\prime}-m)/(m+1)$, $\kappa=\kappa^{\prime}/(m+1)$.  Since
$\kappa + q +1 = (q^{\prime}+\kappa^{\prime}+1)/(m+1)$, the
comparisons between $q$ and $-\kappa -1$ in the following discussion
are equivalent to the comparisons between $q^{\prime}$ and
$-\kappa^{\prime}-1$. At different times, the total energy of the
blast wave may be dominated either by a continuous injection term or
by the initial impulsive term \citep[and references therein]{zm01}.
Which of the two is dominant at a particular observation time $T$
depends both on the values of the two indices ($q$ and $-\kappa -1$),
and on the relative values of ${\cal{L}}_{BZ}$ and $E_{\nu
\overline{\nu}}$.  If $q<-1-\kappa$, the impulsive term always
dominates since the first term is negative.  For an adiabatic blast
wave expanding in a constant density medium (i.e. $m=3$ and
$\kappa=\kappa^{\prime}=0$), this condition yields $q^{\prime}<
-1$. If $q>-1-\kappa$, the continuous term will eventually dominate
(i.e. $E_{\nu \overline{\nu}} \ll {\cal{L}}_{BZ}T_{0}$) over the
second term after a critical time
\begin{equation}
T_{c}=T_{0} \times {\rm max}\left[1,\left((\kappa + q +1){E_{\nu
\overline{\nu}} \over {\cal{L}}_{BZ}T_{0}}\right)^{1/(\kappa+q+1)}\right]. 
\end{equation}
This is the case considered in many pulsar central engines \citep{u92,
u94, dl98}. We then have $T_{c} \approx T_{0}$ and the fireball is
completely analogous to the impulsive regime with $E_{\rm total}
\approx {\cal{L}}_{BZ}T_{0}$. \\

If initially $E_{\nu \overline{\nu}} \gg {\cal{L}}_{BZ}T_0$, the
critical $T_{c}$ after which the continuous injection becomes dominant
could be much longer than $T_{0}$, exerting a noticeable influence on
the GRB afterglow. We expect this to be the case either when the
magnetic energy is below equipartition ($\rho c_{s}^2 > B^{2}/8\pi$)
or when the timescale for the formation of a self--similar solution is
small ($T_{0} \ll T_{c}$). Because of the strong pinch that develops,
a narrow jet that delivers its thrust in a narrow solid angle,
$\Omega_{BZ}$, may be a common ingredient of strong rotating magnetic
fields (see \citet{mku00} for a recent review). We thus generally
expect the magnetic outflow to be much more collimated than that
produced by $\nu \overline{\nu}$ annihilation ($\Omega_{BZ} \ll
\Omega_{\nu \overline{\nu}}$) and the magnetic luminosity to be the
dominant contribution. Even in this case, the impulsive term $E_{\nu
\overline{\nu}}$ may be responsible for either creating a cavity
before the magnetized wind expands or for precursor emission. The
detection of, or strong upper limits on, such features would provide
constraints on the burst progenitor and on magnetar--like central
engine models.

An external shock can occur at much larger radii and over a much
longer timescale than in usual afterglows, if the environment has a
very low density. This may be the case for GRBs arising from compact
binary mergers that are ejected from the host galaxy into an external
medium that is much less dense than the ISM assumed for usual models
(where the particle density is $n \sim 0.1-1$~cm$^{-3}$). It is
commonly assumed that compact mergers occur outside the host galaxies
because of the long inspiral, due to the emission of gravitational
waves. We note that recent calculations of evolutionary tracks of
binary systems containing massive stars show that the merger events
can occur on much shorter timescales and thus still within the host
galaxy, because the initial binary separation is small \citep{bbk02}.

Our simulations would indicate that the central engine survives the
initial, violent event that created it, and that it possesses enough
energy to account for the energetics and durations of short
GRBs. However, they clearly cannot tackle directly other relevant
issues, mainly related to the evolution of the magnetic field, and its
influence on the dynamics. Magnetic instabilities could make the disk
lifetime much shorter by effectively increasing the viscosity. The
amplification of the magnetic field may be self--limiting due to
magnetic stress, which would cause disk flaring. The properties of the
expected variability depend strongly on the details of the
configuration of the disk corona generated by the magnetic field,
which is removed from the disk by flux buoyancy \citep{nara92, tp96,
kl98}.

\acknowledgments

It is a pleasure to acknowledge many helpful conversations with
W. Klu\'{z}niak, D. Lazzati, G. Ogilvie, D. Price, M. J. Rees,
S. Rosswog and V. Usov. We thank the referee for his comments and
suggestions for improvements to the text. WHL thanks the IoA for its
hospitality. Financial support for this work was provided in part by
CONACyT (27987E), PAPIIT (IN-110600), SEP and the Royal Society.

\clearpage

\clearpage

\clearpage

\clearpage

\clearpage

\clearpage

\clearpage

\clearpage

\begin{deluxetable}{ccccccc}
\tablecaption{Initial conditions for the accretion disks.\label{table:ICs}}
\tablewidth{0pt}
\tablehead{
\colhead{Run} & \colhead{$q_{b}$\tablenotemark{a}} & \colhead{$\Gamma$} & \colhead{$\alpha$} & \colhead{$M_{\rm BH}/M_{\sun}$\tablenotemark{b}} & \colhead{$M_{disk}/M_{\sun}$\tablenotemark{b}} &  \colhead{$N$\tablenotemark{c}} 
}
\startdata
A & 0.31 & 2.0 & 0.1 & 5.57 & 0.266 & 21,609 \\
B & 0.31 & 4/3 & 0.1 & 5.57 & 0.266 & 21,609 \\
C & 0.50 & 2.0 & 0.1 & 3.85 & 0.308 & 19,772 \\
D & 0.50 & 4/3 & 0.1 & 3.85 & 0.308 & 19,772 \\
E & 0.50 & 4/3 & 0.1 & 3.85 & 0.308 & 79,489 \\
F & 0.50 & 4/3 & 0.01 & 3.85 & 0.308 & 19,772 \\
G & 0.50 & 4/3 & 0.01 & 3.85 & 0.308 & 39,658 \\
\enddata
\tablenotetext{a}{Mass ratio $q_{b}=M_{\rm NS}/M_{\rm BH}$ of the original binary system used in the three--dimensional simulation.}
\tablenotetext{b}{Values at the start of the two--dimensional calculation of the accretion disk evolution.}
\tablenotetext{c}{Number of SPH particles at the start of the two--dimensional calculation.}

\end{deluxetable}

\clearpage

\begin{deluxetable}{cccccccc}
\tabletypesize{\scriptsize}
\tablecaption{Accretion disk parameters during the dynamical evolution.\label{table:evol}}
\tablewidth{0pt}
\tablehead{
\colhead{Run} & \colhead{$\dot{M}/(M_{\sun}/s)$\tablenotemark{a}} & \colhead{$\rho c_{s}^{2}/(\mbox{erg}~\mbox{cm}^{-3})$\tablenotemark{a}} & \colhead{$L_{\nu}(\mbox{erg}~\mbox{s}^{-1})$\tablenotemark{a}} & \colhead{$E_{\nu}(\mbox{erg})$\tablenotemark{b}}  & \colhead{$L_{BZ}(\mbox{erg}~\mbox{s}^{-1})$\tablenotemark{a}} & \colhead{$E_{BZ}(\mbox{erg})$\tablenotemark{b}}  & \colhead{$M_{disk}/M_{\sun}$\tablenotemark{b}} 
}
\startdata
A & 0.33 & 1.8$\cdot10^{30}$ & 3.25$\cdot10^{52}$ & 1.38$\cdot10^{52}$  & 1.00$\cdot10^{51}$ & 2.94$\cdot10^{50}$ & 0.128  \\
B & 0.31 & 2.1$\cdot10^{30}$ & 3.61$\cdot10^{52}$ & 1.08$\cdot10^{52}$  & 1.75$\cdot10^{51}$ & 4.05$\cdot10^{50}$ & 0.151  \\
C & 0.36 & 3.5$\cdot10^{30}$ & 4.00$\cdot10^{52}$ & 1.79$\cdot10^{52}$  & 2.00$\cdot10^{51}$ & 6.47$\cdot10^{50}$ & 0.137  \\
D & 0.48 & 5.0$\cdot10^{30}$ & 5.00$\cdot10^{52}$ & 1.52$\cdot10^{52}$  & 2.50$\cdot10^{51}$ & 6.25$\cdot10^{50}$ & 0.158  \\
E & 0.47 & 5.3$\cdot10^{30}$ & 4.65$\cdot10^{52}$ & 1.57$\cdot10^{52}$  & 2.95$\cdot10^{51}$ & 7.41$\cdot10^{50}$ & 0.160  \\
F & 0.13 & 12.5$\cdot10^{30}$ & 1.86$\cdot10^{52}$ & 0.44$\cdot10^{52}$  & 6.00$\cdot10^{51}$ & 1.17$\cdot10^{51}$ & 0.254 \\ 
G & 0.15 & 12.0$\cdot10^{30}$ & 1.84$\cdot10^{52}$ & 0.44$\cdot10^{52}$  & 5.65$\cdot10^{51}$ & 1.15$\cdot10^{51}$ & 0.255  \\
\enddata
\tablenotetext{a}{Values are given at $t=0.1$~s.}
\tablenotetext{b}{Values are given at $t=0.2$~s.}

\end{deluxetable}

\end{document}